\documentclass[sigconf,natbib = true]{acmart}
\pdfoutput=1

\copyrightyear{2019} 
\acmYear{2019} 
\setcopyright{acmlicensed}
\acmConference[ACSAC '19]{2019 Annual Computer Security Applications Conference}{Dec. 9--13, 2019}{San Juan, PR, USA}
\acmBooktitle{2019 Annual Computer Security Applications Conference (ACSAC '19), Dec. 9--13, 2019, San Juan, PR, USA}
\acmPrice{15.00}
\acmDOI{10.1145/3359789.3359806}
\acmISBN{978-1-4503-7628-0/19/12}

\DeclareUrlCommand\Code{\urlstyle{rm}}
\expandafter\def\expandafter\UrlBreaks\expandafter{\UrlBreaks  
\do\/\do\a\do\b\do\c\do\d\do\e\do\f\do\g\do\h\do\i\do\j\do\k
\do\l\do\m\do\n\do\o\do\p\do\q\do\r\do\s\do\t\do\u\do\v
\do\w\do\x\do\y\do\z
\do\A\do\B\do\C\do\D\do\E\do\F\do\G\do\H\do\I\do\J\do\K
\do\L\do\M\do\N\do\O\do\P\do\Q\do\R\do\S\do\T\do\U\do\V
\do\W\do\X\do\Y\do\Z}

\definecolor{darkspringgreen}{rgb}{0.09, 0.45, 0.27}
\definecolor{dartmouthgreen}{rgb}{0.05, 0.5, 0.06}

\usepackage{tikz}
\usetikzlibrary{intersections}	

\usetikzlibrary{matrix}
\usepackage{pifont}
\definecolor{light-gray}{gray}{0.80}
\definecolor{darkgray}{rgb}{0.66, 0.66, 0.66}
\definecolor{gray}{rgb}{0.5, 0.5, 0.5}

\usepackage{xspace}
\newcommand{\sysname}{{\small\textsc{LLVM-CFI}}\xspace}

\makeatother

\begin{document}
\sloppy

\pagestyle{plain} 



\title{Analyzing Control Flow Integrity with LLVM-CFI}

\author{Paul Muntean}
\orcid{https://orcid.org/0000-0002-2462-7612}
\affiliation{%
\institution{Technical University of Munich}
}
\email{paul.muntean@sec.in.tum.de}

\author{Matthias Neumayer}
\affiliation{%
\institution{Technical University of Munich}
}
\email{matthias.neumayer@tum.de}

\author{Zhiqiang Lin}
\affiliation{%
\institution{The Ohio State University}
}
\email{zlin@cse.ohio-state.edu}

\author{Gang Tan}
\affiliation{%
\institution{Penn State University}
}
\email{gtan@psu.edu}

\author{Jens Grossklags}
\affiliation{%
\institution{Technical University of Munich}
}
\email{jens.grossklags@in.tum.de}

\author{Claudia Eckert}
\affiliation{%
\institution{Technical University of Munich}
}
\email{claudia.eckert@sec.in.tum.de}

\begin{abstract}
Control-flow hijacking attacks are used to  perform malicious computations.
Current solutions for assessing the attack surface after a \textit{control flow integrity} (CFI) policy was applied can measure
only indirect transfer averages in the best case without providing any insights w.r.t. the absolute calltarget reduction per
callsite, and gadget availability. Further, tool comparison is underdeveloped or not possible at all.
 CFI has proven to be one of the most promising protections against control flow hijacking attacks, thus many efforts have been made to improve CFI in various ways. However, there is a lack of systematic assessment of existing CFI protections. 
%
%

In this paper, we present \sysname, a static source code analysis framework for analyzing state-of-the-art static CFI protections based on the Clang/LLVM compiler framework. 
\sysname works by precisely modeling a CFI policy and then evaluating it within a unified approach. \sysname helps determine the level of security offered by different CFI protections, after the CFI protections were deployed, thus providing an important step towards exploit creation/prevention and
stronger defenses. We have used \sysname to assess eight state-of-the-art static CFI defenses on real-world programs such as Google Chrome and Apache Httpd.
\sysname provides a precise analysis of the residual attack surfaces, and accordingly ranks CFI policies against each other. \sysname also successfully paves the way towards construction of COOP-like code reuse attacks and  elimination of the remaining attack surface by disclosing protected calltargets under eight restrictive CFI policies.

\end{abstract}


\ccsdesc[500]{Security and privacy~Systems security}
\ccsdesc[300]{Software and application security}

\keywords{Clang, LLVM, control flow integrity, computer systems, defense.}

\maketitle

\section{Introduction}
\label{Introduction}  
Ever since the first Return Oriented Programming (ROP) attacks \cite{ret-into-lib, krahmer, rop:shacham}, the cat and mouse game between defenders and attackers has initiated a plethora of research.
As defenses improved over time, the attacks progressed with them, as pointed out by Carlini \textit{et al.} \cite{rop:carlini}. While defenders followed several lines of research when building defenses: control flow integrity \cite{zhang:vtrust, ropecker, perinput:niu, marx:tool, vci:asiaccs, vtint:zhang, vtv:tice, ivt, shrinkwrap, mcfi:niu, veen:typearmor,ptcfi:2017,paul:castsan, paul:raid}, 
binary re-randomization \cite{shuffler},
information hiding \cite{oxymoron}, and 
code pointer integrity \cite{cpi}, 
the attacks kept up the pace and got more sophisticated \cite{coop, coop:loop:oriented, subversive-c:lettner, blue:lotus, trap:crane}.

In principle, even with the myriad of currently available CFI defenses, performing exploits is still possible.
This holds even in the presence of hypothetically perfect CFI~\cite{rop:carlini}. For this reason, in this work, we aim to answer the question of how secure are programs which are protected by CFI defenses. Even after CFI defenses are in place, attackers could search the program for gadgets that are allowed (for example, pass under the radar due to imprecision) by CFI defenses 
to conduct \textit{Code Reuse Attacks} (CRAs); see \cite{carlini:bending, coop}. As such, these attacks become highly program-dependent,
and the applied CFI policies make reasoning about security harder. 
The attacker/analyst is thus confronted with the challenge of searching (manually or automatically) the protected program's binary or source code for gadgets which remain useful after CFI defenses have been deployed. As a result, there is a growing demand for defense-aware attack analysis tools, which assist security analysts when analyzing CFI defenses.

To the best of our knowledge, there are neither tools for statically modeling and comparing static CFI defenses against each other, nor static CRA crafting tools which are aware of a set of applied defenses.
Existing tools, including static pattern-based gadget searching tools \cite{ropdefender, multiarchitecture:wollgast} and dynamic attack construction tools \cite{newton, revery, jujutsu, losing:control:conti, bop}, all lack deeper knowledge of the protected program. 
As such, they can find CRA gadgets, but cannot determine if the gadgets are usable after a defense was deployed.
Consequently, with each applied defense, a more capable assessment tool needs ideally to: 
(1) model the defense as precisely as possible, 
(2) use program metadata in order not to solely rely on runtime memory constraints,
(3) use precise semantic knowledge of the protected program, and
(4) provide absolute analysis numbers w.r.t. the remaining attack surface after a defense was deployed. This allows an analyst to provide precise and reproducible measurements, to decide which CFI defense is better suited for a given situation, and to defend against or craft CRAs by searching available gadgets. Lastly, none of the existing static and dynamic program analysis tools can be used to compare static CFI defenses against each other w.r.t. the offered protection level after deployment.

In this paper, we present \sysname, which to the best of our knowledge, is the first static Clang/LLVM based compiler framework used for modeling and analyzing static state-of-the-art CFI defenses. \sysname can determine the security level these protection techniques offer as well as the remaining attack surface after such a defense was deployed. \sysname automates one step of COOP-like attacks by finding protected targets towards which program control-flow can transfer. As such, \sysname provides the first step towards searching for COOP-like gadgets, but its main purpose is to evaluate static CFI policies against each other. Further, \sysname cannot build CRAs automatically but rather assist on how one could go about when constructing CRAs.
We envisage \sysname to be used as a tool to analyze conceptual or deployed CFI defenses by either an 
analyst---to better compare existing CFI defenses against each other---or by an attacker (\textit{e.g.,} red team attacker)---to help craft attacks which can bypass existing in-place CFI defenses.

\sysname is a {\em unified framework} to evaluate different CFI defenses, enabling a head-to-head comparison. To achieve this, we introduce a new CFI defense analysis metric dubbed calltarget reduction (\textit{CTR}), which tells precisely, without averaging the results, how many calltargets are still available after a CFI defense was enforced. \sysname is capable to analyze CFI policies w.r.t. several metrics and thus compare CFI policies w.r.t. different aspects. $CTR$ is one example metric along other three metrics presented in this paper. \looseness=-1

Further, we are particularly interested in calltarget reduction analysis as this is the most used metric (see AIR~\cite{mingwei:sekar}, fAIR~\cite{vtv:tice}, and AIA~\cite{aia} --- however, these metrics average the results) to compare CFI defenses against each other. At the time of writing this paper, none of the existing CFI metrics can tell how secure a program is after a certain CFI policy was applied; as such, we do not claim that by using our $CTR$ metric we can provide more security guarantees than other metrics, but rather $CTR$ provides absolute values rather than averaging them.
Even though the calltargets could be used or not during an attack, we opt in this work to not further investigate this avenue as the usability of a target depends on the type of CRA performed. Instead, for each protected callsite, we show additional calltarget features (see Section \ref{rq7}), such that the analyst could with ease figure out if the targets are usable for a particular attack.

By using different compilers, compiler flag settings or OSs, the results of CFI policy analysis could not be comparable against each other. For this reason, \sysname relies on the insight that, by precisely modeling a CFI defense into a comprehensive policy, the introduced constraints on callsites and calltargets can be assessed during program compile time, by an unified analysis framework. Further, in order to support this task, \sysname provides a set of program {\em expressive primitives}, which help to characterize a wide range of static CFI policies. For example, \sysname offers static primitives related to variable types, class hierarchies, virtual table layouts and function signatures. These primitives can be used as building blocks to model a wide range of CFI policies. Further, \sysname provides the legitimate calltarget set for each callsite under different CFI defenses. 
This set can be evaluated by a lower and upper target bound. The lower bound is represented by the set of all calltargets which are, according to the analyzed policy (\textit{e.g.,} sub-hierarchy policy \cite{ivt}), legitimate to be called by a protected callsite during benign program execution. Accordingly, the upper bound is represented by the set of all calltargets that can be called (both legitimately or not) by a protected callsite during benign program execution (\textit{e.g.,} all virtual tables policy \cite{vtint:zhang}).
Further, \sysname paves the way towards automated control-flow hijacking attack construction, \textit{e.g.,} the control flow bending attack, see Carlini \textit{et al.} \cite{carlini:bending}, or to refine the analysis of existing attack construction or defense tools. \looseness=-1

\sysname analyzes statically the CFI defenses, as these are more commonly deployed against control-flow hijacking attacks than dynamic defenses. Further, \sysname focuses on source code (LLVM's IR and Clang metadata is pushed into compiler's LTO phase and analyzed) rather than on binary code, as comparing various static CFI defenses against each other is feasible only in this way. Moreover, the binary CFI policy implementations can be more precisely expressed with source code at hand.
Therefore, we opt not to analyze the machine code of the protected programs as source code provides more semantics and precision w.r.t. the constraints imposed by each CFI defense. Thus, \sysname models static CFI policies for binaries more precisely than the binary instrumentation tools which were used to enforce them as these operated mostly on the binary only. Lastly, \sysname models source code based CFI policies as precise as the original policies as these were implemented either atop Clang/LLVM or GCC compilers.

We evaluated \sysname with real-world programs including Google's Chrome, NodeJS, etc., and with eight state-of-the-art static CFI policies which were previously thoroughly discussed and clarified with their original authors w.r.t. how these were modeled within \sysname. We selected eight representative binary and source code based CFI policies based on the criteria that the policies should be static, state-of-the-art, and available as open source (published). Further, we are aware that there are other CFI policies which cannot currently be modeled with \sysname. For this reason, in this paper, we do not aim to give a full presentation of which CFI policies can be modeled with \sysname.

\sysname can help the assessment of CFI defenses and it can help at finding gadgets, even with highly restrictive CFI defenses deployed. Further, we demonstrate how \sysname can be utilized to pave the way towards automated CRA construction. We also show how it can be effectively used to empirically measure the real attack surface reduction after a certain static CFI defense policy was used to harden a program's binary. 

Comparing binary and software based policy results against each other is out of scope of this paper. Rather, the goal of our tool is to show how the analyzed CFI policies compare against each other and to provide insights on their precision and  benefits.
Applications of \sysname go beyond a CFI defense assessment framework, and we envision \sysname as a tool for defenders and software developers to highlight the residual attack surface of a program. As such, a programmer can evaluate if a bug at a certain program
location enables a practical CRA.

In summary, our contributions include the following:
\begin{itemize}

\item We implement \sysname\footnote{The source code is available at \url{https://github.com/TeamVault/LLVM-CFI.git}}, a novel framework for empirically analyzing and comparing CFI defenses against each other. 

\item We introduce our comprehensive formalization framework for formalizing static state-of-the-art binary and source code based CFI defenses.

\item We show evaluation results based on real-world programs by comparing existing static CFI defenses against each other.


\item We present an attacker model that is powerful and drastically lowers the bar for performing attacks against state-of-the-art CFI defenses with \sysname at hand. 
\end{itemize}

The remainder of this paper is organized as follows.
Section \ref{Background} contains the brief overview of the required technical background knowledge.
Section \ref{Overview} describes the design of \sysname, and
Section \ref{implementation} presents implementation details of \sysname, while Section \ref{Evaluation} contains the evaluation of \sysname.
Section \ref{Related Work} highlights related work 
and Section~\ref{Conclusions} offers concluding remarks.

\section{Background} 
\label{Background} 

\textit{\textbf{Control Flow Integrity.}}
Control-Flow Integrity (CFI) \cite{abadi:cfi, abadi:cfi2} is a state-of-the-art technique used successfully along other techniques to protect forward and backward edges against program control-flow hijacking based attacks.
CFI is used to mitigate CRAs by, for example, pre-pending an indirect callsite with runtime checks that make sure only legal calltargets are allowed by an as-precisely-as-possible  computed control flow graph (CFG). 

\medskip\noindent\textit{\textbf{Protection Schemes.}}
Alias analysis in binary programs is undecidable \cite{alias:analysis}. For this reason, when protecting CFG forward edges, defenders focus on using other program
primitives to enforce a precise CFG during runtime.
These primitives are most commonly represented by the program's: class hierarchy~\cite{shrinkwrap}, virtual table layouts \cite{vfuard:aravind}, 
reconstructed-class hierarchies from binary code~\cite{marx:tool}, binary function types \cite{veen:typearmor} (callsite/calltarget parameter count matching), etc.
They are used to enforce a CFG which is as close as possible to the original CFG being best described by the program control flow execution. 
Note that state-of-the-art CFI solutions use either static or dynamic information for determining legal calltargets.


\medskip\noindent\textit{\textbf{Static Information.}}
CFI defenses which are based on static information allow, for example, callsites to target:
(1) function entry points, \textit{e.g.,} \cite{mingwei:sekar},
map callsite types to target function types by creating a mask which enforces that the number of provided parameters (up to six) has to be higher than the number of consumed parameters, \textit{e.g.,} \cite{veen:typearmor},
(2) a rebuilt-class hierarchy (no root node(s) and the edges are not oriented) can be recuperated from the binary and enforced, \textit{e.g.,} \cite{marx:tool},
(3) all virtual tables that can be recuperated and enforced, \textit{e.g.,} \cite{vfuard:aravind},
only certain virtual table entries are allowed, \textit{e.g.,} \cite{zhang:vtrust} based on a precise function type mapping, and
(4) sub-class hierarchies are enforced, \textit{e.g.,} \cite{vtv:tice, shrinkwrap, ivt}. Thus, in order for such techniques to work, program metadata should be available or it should be possible through program analysis to reconstruct it.


\section{Design} 
\label{Overview} 
\subsection{Overview}

\sysname is designed to assist an analyst 
evaluating the attack surface after different types of static CFI defenses were applied and to pave the way towards automated detection of
gadgets.
To achieve this, \sysname applies a static black box strategy in order to statically retrieve the set of attacker-controllable forward control flow graph (CFG) edges. The forward-vulnerable CFG edges are expressed as a callsite with a variable number of possible target functions. Further, these CFG edges can be reused by an attacker to call arbitrary functions via arbitrary read or write primitives. To call such series of arbitrary functions, an attacker can chain a number of edges together by dispatching fake objects
contained in a vector. See, for example, the COOP \cite{coop} attack which is based on a dispatcher gadget used to call other gadgets through a single loop iteration. The COOP attack uses gadgets which are represented by whole virtual functions.

\sysname supports a wide range of code reuse defenses based on user-defined policies, which are composed of constraints about the set of possible calltargets allowed by a particular applied CFI defense.
The main idea behind \sysname is to compile the target program with different types of CFI policies 
and get the allowed target set per callsite for each constraint configuration. 
Note that we assume the program was compiled with the same compiler as the one on which \sysname is based.
Moreover, \sysname's policies are reusable and extensible; they model security invariants of important CRA defenses. Essentially, under these 
constraints, virtual pointers at callsites can be corrupted to call any function in the program. Thus, in this paper, we focus 
on the possibility to bend \cite{carlini:bending}
a pointer to the callsite's legitimate targets. Further, we assume that large programs contain enough gadgets for successfully performing CRAs. Bending assumes that it is possible for an attacker to reuse protected gadgets during an attack making the applied defense of questionable benefit.

\begin{figure}[ht]
\centering
  \vspace{.4cm}
   \centering
   \includegraphics[width = 0.4\textwidth, bb = 0 0 226 162]{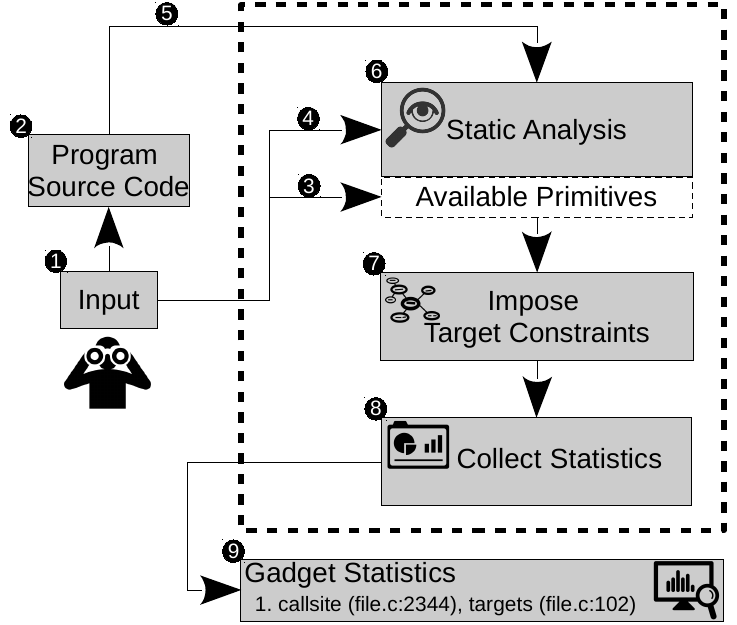}
    \caption{Design of \sysname.}
    \label{Design of CFI-Assessor.}
\end{figure}
 
\autoref{Design of CFI-Assessor.} depicts the main components of \sysname and the workflow used to analyze the source code of a potentially vulnerable program in order to determine CRA statistics, as follows.
To use \sysname the analyst has to provide as input \ding{182} to \sysname a program's source code \ding{183}. 
In addition, the analyst needs to choose the desired defense policies. A subset of defense policies can be selected by switching on flags inside the \sysname source code, which can also be implemented as compiler flags, if desired. 
Fundamentally, the analysis performed by \sysname is dependent on the implemented defense models and on the available primitives. Therefore, the analyst can choose from the 8 policies currently available in the \sysname tool. In case a desired defense is not available, the analyst can extend the list of primitives \ding{184}, and model his defense as a policy (set of constraints) in the analysis component of \sysname~\ding{185}. 
In order to do this, he needs to know about the analysis internals of \sysname. 
After selecting/modeling a defense, the analyst forwards the application's source code \ding{186}  to \sysname which will analyze it using its static analysis component.
During static analysis \ding{187} the previously selected defense will be applied when compiling the program source code. As the analysis is performed, each callsite is constrained with only the legitimate calltargets. Note that both the protected callsites as well as the
legitimate calltargets per callsite are dependent on the currently selected defense model.
The result of the analysis contains information about the residual target set for each individual callsite after a CFI policy was assessed \ding{188}. This list contains 
a set of gadgets (callsites + calltargets) that can, given a certain defense model, be used to bend the control flow of the application. 
These target constraints are collected and clustered in the statistics collection component of \sysname~\ding{189}.
Finally, at the end of the gadget collection phase, a list of calltargets containing potential usable gadgets statistics \ding{190} based on the currently applied defense(s) will be reported. 

\subsection{Analysis Primitives}
\label{Available Primitives}
\sysname provides the following program primitives, which are either collected or constructed during program compile time. 
These primitives are used by \sysname to implement static CFI policies and to perform calltarget constraint analysis. 
Briefly, the currently available primitives are as follows:

\textbf{Virtual table hierarchy} (see \cite{shrinkwrap} for a more detailed definition) allows performing virtual table inheritance analysis of only virtual classes as only 
 these have virtual tables. Finally, a class is virtual if it defines or inherits at least one virtual function.
 
\textbf{Vtable set} is a set of vtables corresponding to a single program class.
 This set is useful to derive the legitimate set of calltargets for a particular 
 callsite. The set of calltargets is determined by using the class inheritance relations contained inside a program.
 
\textbf{Class hierarchy} (see Tip \textit{et al.} \cite{class:slicing} and Rossie \textit{et al.} \cite{rossie:friedman} for a more formal definition) can be represented as a class hierarchy graph with the goal to model inheritance relations between classes. Note that a real-world program can have multiple class hierarchies (\textit{e.g.,} Chrome, Google's Web browser). Note that the difference between virtual table hierarchy and class hierarchy is that the class hierarchy contains both virtual and non-virtual classes, 
whereas the virtual table hierarchy can only be used to reason about the inheritance relations between virtual classes.
  
\textbf{Virtual table entries} allow \sysname to analyze the number of entries in each virtual table with the possibility to differentiate between
 virtual function entries, offsets in vtables, and thunks.

\textbf{Vtable type} is determined by the name of the vtable root for a given vtable. A vtable root is the last derived 
 vtable contained in the vtable hierarchy.
 
\textbf{Callsites} are used by \sysname to distinguish between direct and
 indirect (object-based dispatch and function-pointer based indirect transfers) callsites.

\textbf{Indirect callsites} are based on: (1) object dispatches or (2) function pointer based calls.
 Based on these primitives, \sysname can establish different types of relations between 
 callsites and calltargets (\textit{i.e.,} virtual functions). At the same time, we note that it is possible to derive 
 backwards relationships from calltargets to legitimate callsites based on this primitive.
 
\textbf{Callsite function types} allow \sysname to precisely determine the number and the type of the provided parameter by a callsite.
 As such, a precise mapping between callsites and calltargets is possible.

\textbf{Function types} allow \sysname to precisely determine the number 
 of parameters, their primitive types and return type value for a given function. 
This way, \sysname can generate a precise mapping between compatible
 calltargets and callsites.

These primitives can be used as building blocks during the various analyses that \sysname
can perform in order to derive precise measurements and a thorough assessment of a modeled static CFI policy.
We note that in order to model other CFI defenses, other (currently not available) simple or aggregated
analysis primitives may need to be added inside \sysname. \looseness=-1

\subsection{Constraints}
\label{Target Constraints Analysis}

The basic concept of any CRA is to divert the intended control flow of a program by using arbitrary memory write and read primitives.
As such, the result of such a corruption is to bend~\cite{carlini:bending} the control flow, such that it no longer points to  
the intended (legitimate) calltarget set. This means that 
the attacker can point to any memory address in the program. While this type of attack is still possible, we want to highlight another type of CRA
in which the attacker uses the intended/legitimate per callsite target set. That is, the attacker calls inside this set and performs her malicious behavior by reusing calltargets which are protected, yet usable during an attack. As previously observed by others \cite{newton}, CRA defenses try to mitigate this by mainly relying on one or two dimensions at a time, as follows:

 \textbf{{Write constraints}} limit the attacker's capabilities to corrupt writable memory. If there is no defense in place, 
 the attacker can essentially corrupt: pointers to data, non-pointer values such as strings, and pointers to code (\textit{i.e.,} 
 function pointers). In this paper, we do not investigate these types of defenses as these were already addressed in detail by Veen \textit{et al.} \cite{newton}. 
 Instead, we focus on target constraints as these represent a big class of defenses which in our assessment need a separate and detailed  treatment. This obviously does not mean that our analysis results cannot be used in conjunction with dynamic write constraint 
 assessing tools. Rather, our results represent a common ground truth on which runtime assessing tools can build their gadget detection analysis.
 
 {\textbf{Target constraints}} restrict the legitimate calltarget set for a callsite which can be controlled by an attacker.
 With no target constraints in place, the target set for each callsite is represented by all functions located in the program and any linked shared library.
 The key idea is to reduce the wiggle room for the attacker such that he cannot target random callsites. 
 As most of these defenses impose a one-to-$N$ mapping, an attacker being aware of said mapping could corrupt the pointer at the callsite to bend \cite{carlini:bending}
 the control flow to legitimate targets
 in an illegitate
 order to achieve her malicious goals. Thus, all static CFI defenses impose target constraints.

\medskip\noindent\textbf{Static Analysis.} 
\sysname is based on the static analysis of the program which is represented in LLVM's intermediate representation (IR). The analysis is performed 
during link time optimization (LTO) inside the LLVM \cite{llvm:framework} compiler framework to detect callsites and legitimate callees under the currently analyzed CFI defense. 
\sysname uses the currently available primitives and the implemented defenses to impose target constraints for each callsite individually. 
Currently, eight defenses are supported, see Section \ref{implementation}, but this list can easily be extended since 
all defenses are based on similar mechanisms which are assessable during a whole program analysis. 

\medskip\noindent\textbf{Generic Target Constraints.} 
As mentioned above, \sysname can be used to impose existing generic calltarget constraints (defenses) based on class hierarchy relations and 
callsites and calltarget type matching with different levels of precision depending on the currently modeled CFI defense. 
Further, \sysname allows 
extending and combining existing policies or applying them concurrently.


\subsection{Describing CFI Defenses} 
\label{Mapping Defenses Into CFI-Assessor} 

In this section, we present our CFI defense formalization framework and how this was used to formalize eight static CFI defenses inside \sysname. 
These defenses are stemming from published research papers and are used to constrain forward edge program control flow transfers to point to only legitimate calltargets.
Note that each CFI defense description is an idealized representation and very close to how the original CFI defense policy was implemented in each tool. Further, in case of the binary based CFI defenses, the CFI defense descriptions are more precise than their implementations in the respective tools.
Lastly, note that each modeled defense was previously discussed with the original authors and only after the authors agreed with these descriptions we modeled them into \sysname. Next, we give the formal definitions of 
each of the CFI defenses as these were modeled inside \sysname and the descriptions of the performed analyses.

\begin{table}[ht!]
\centering
\resizebox{.92\columnwidth}{!}{%
\begin{tabular}{l|l}
 \textbf{Symbol} &  \ \ \ \ \ \ \ \ \ \ \ \ \ \ \ \ \ \ \ \ \ \ \textbf{Description}  \\\hline
$P$ & the analyzed program \\
$Cs$ & set of all indirect callsites of $P$ \\
$Cs_{virt}$ & set of $P$ virtual callsites \\
$V$ & all virtual func. contained in a virtual table hierarchy \\
$V_{sub}$ & a virtual table sub-hierarchy \\
$v_t$ & a virtual table \\
$v_e$ & a virtual table entry (virtual function) \\
$vc_s$ & a virtual callsite \\
$nv_f$ & a non-virtual function \\
$v_f$ & a virtual function (virtual table entry) \\
$C$ & a class hierarchy contained in $P$ \\
$C_{sub}$ & a class sub-hierarchy contained in $P$\\
$c_s$ & an indirect callsite \\
$nt_{pcs}$ & callsite's number and type of parameters \\
$nt_{pct}$ & calltarget's number and type of parameters \\
$F$ & set of all virtual and non-virtual functions in $P$ \\
$F_{virt}$ & set of all virtual functions in $P$ \\
$S$ & set of function signatures \\
$M$ & calltarget matching set based on the policy rules \\
\end{tabular}
}
\caption{Symbol descriptions.}
\label{notation}
\end{table}
\medskip\noindent\textbf{Notation.} \autoref{notation} depicts the set of symbols used by \sysname to model CFI defenses.
Note that $M$ is determined by applying all rules defined by a CFI defense and represents, at the same time, the matching criteria for each policy. This means that \sysname increments the count of its analysis by one when such a match is found. \looseness=-1

\medskip\noindent\textbf{Bin Types. (TypeArmor) \cite{veen:typearmor}} 
  \label{typearmor:policy} 
  We formalize this CFI policy $\Psi$ as the tuple   
  $\big \langle Cs, F, V, M \big \rangle$ where the relations hold:
  (1) $V \subseteq F$,
  (2) $v_e \in V$,
  (3) $nv_f \in F$,
  (4) $c_s \in C$, and
  (5) $M \subseteq Cs \times V \times F$.
  
  \textit{\sysname's Analysis.} For each indirect callsite $c_s$ (1) count the total number of virtual table entries $v_e$ which reside in each virtual table hierarchy $V$ contained in program $P$, and also, (2) count the number of non-virtual functions $nv_f$ residing in $F$, which need at most as many function parameters as provided by the callsite and up to six parameters. 
  Further, if $F$ contains multiple distinct virtual table hierarchies (islands) then continue to count them too and take them also into consideration for a particular callsite. 
  An island is a virtual table hierarchy which has no father-child relation to another virtual table hierarchy contained in the program $P$.
 
\medskip\noindent  \textbf{Safe src types. (Safe IFCC) \cite{vtv:tice}}
    We formalize this CFI policy $\Psi$ as the tuple   
  $\big \langle Cs, F, F_{virt} S, M \big \rangle$ where the relations hold:
  (1) $V \subseteq F$,
  (2) $v_f \in F_{virt}$,
  (3) $nv_f \in F$,
  (4) $nt_{pcs} \in S$,
  (5) $nt_{pct} \in S$,
  (6) $f_{rt} \in S$,
  (7) $c_s \in Cs$, and
  (8) $M \subseteq Cs \times F \times S$.
  
    \textit{\sysname's Analysis.} {For each indirect callsite $c_s$ count the number of virtual functions $v_f$ and non-virtual functions $nv_f$ located in the program $P$ for which the number and type of parameters required by the calltarget $nt_pct$ matches the number and type of parameters provided at the callsite $nt_pcs$. 
 The function return type $f_{rt}$ of the matching function is not taken into consideration. 
 All parameter pointer types are considered 
 interchangeable, \textit{e.g.,} \textbf{int*} and \textbf{void*} pointers are considered interchangeable.}

\medskip\noindent  \textbf{Src types. (IFCC/MCFI) \cite{mcfi:niu}} 
   We formalize this CFI policy $\Psi$ as the tuple   
  $\big \langle Cs, F, F_{virt} S, M \big \rangle$ where the relations hold:
  (1) $V \subseteq F$,
  (2) $v_f \in F_{virt}$,
  (3) $nv_f \in F$,
  (4) $nt_{pcs} \in S$,
  (5) $nt_{pct} \in S$,
  (6) $f_{rt} \in S$, 
  (7) $c_s \in Cs$, and 
  (8) $M \subseteq Cs \times F \times S$.
  
    \textit{\sysname's Analysis.} {For each indirect callsite $c_s$ count the number of virtual functions and non-virtual functions located in the program $F$
 for which the number and type of parameters required at the calltarget $nt_{pct}$ matches the number and type of arguments provided by the callsite $nt_{pcs}$. 
 The return type of the matching function is ignored. Compared to \textit{Safe src types}, this policy distinguishes between different pointer types. This means that these are not interchangeable and that the function signatures are more strict.
 Neither the return value of the matching function nor the name of the function are taken into consideration.}
 
\medskip\noindent \textbf{Strict src types. (vTrust) \cite{zhang:vtrust}} 
  We formalize this CFI policy $\Psi$ as the tuple   
  $\big \langle Cs, V, F, F_{virt}, S, M \big \rangle$ where the relations hold:
  (1) $V \subseteq F$,
  (2) $v_f \in F_{virt}$,
  (3) $ntf_{pcs} \in S$,
  (4) $f_{rt} \in S$, 
  (5) $c_s \in Cs$, and 
  (6) $M \subseteq Cs \times S \times F_{virt} \times V$.
  
  \textit{\sysname's Analysis.} {For each indirect callsite $c_s$ compute the function signature of the function called at this particular callsite using the number of parameters, their types, and the name of the function $ntf_{pcs}$ (the literal name used in \verb!C/C++! without any class information attached). Match this function type identifier with each virtual function $v_f$
contained in each virtual table hierarchy $V$ of $P$. The name of the function is taken into consideration when building the hash,
but not the function return type $f_{rt}$, as this can be polymorphic. We have a match when the signature of a function called by a callsite matches the signature of a virtual function $v_f$.}
 
\medskip\noindent \textbf{All vtables. (vTint) \cite{vtint:zhang}} 
  We formalize this CFI policy $\Psi$ as the tuple   
  $\big \langle P, Cs, F_{virt}, V, M \big \rangle$ where the relations hold:
  (1) $V \subseteq F$,
  (2) $v_e \in V$,
  (3) $v_f \in F_{virt}$,
  (4) $c_s \in Cs$, and 
  (5) $M \subseteq Cs \times V$.
  
   \textit{\sysname's Analysis.} {For each indirect callsite $cs$ count each virtual function $v_f$ corresponding to a virtual table entry $v_e$ contained in each virtual table present in the program $P$.}
 
\medskip\noindent \textbf{vTable hierarchy/island. (Marx) \cite{marx:tool}} 
  We formalize this CFI policy $\Psi$ as the tuple   
  $\big \langle P, F_{virt}, C, Cs, V, M \big \rangle$ where the relations hold:
  (1) $V \subseteq F$,
  (2) $v_e \in V$,
  (3) $v_f \in F_{virt}$,
  (4) $v_t \in V$,
  (5) $V \in C$,
  (6) $C \in P$,
  (7) $c_s \in Cs$, and
  (8) $M \subseteq Cs \times V \times C$.
  
  \textit{\sysname's Analysis.} {For each indirect callsite $c_s$ count each virtual function $v_f$ corresponding to each virtual table $v_t$ entry $v_e$ having the same index in 
 the virtual table as the index determined at the callsite $c_s$ by Marx.
 Perform this matching for each virtual table $v_t$ where the index matches the index determined at the callsite $c_s$ and which is located in the class hierarchy $C$ which contains the class type of the dispatched object.
 Note that abstract classes are not taken in consideration within this policy, this can be recognized though by virtual tables having pure virtual function entries.}  \looseness=-1

\medskip\noindent  \textbf{Sub-hierarchy. (VTV) \cite{vtv:tice}} 
   We formalize this CFI policy $\Psi$ as the tuple   
  $\big \langle P, F_{virt}, C, C_{sub}, V, M \big \rangle$ where the relations hold:
  (1) $v_t \in V$,
  (2) $V \subseteq C$,
  (3) $C \subseteq P$,
  (4) $C_{sub} \in C$,
  (5) $v_f \in F_{virt}$,
  (6) $vc_s \in P$, and 
  (7) $M \subseteq Cs_{virt} \times C_{sub} \times V \times F_{virt}$.
  
    \textit{\sysname's Analysis.} {For each virtual callsite $vc_s$ build the class sub-hierarchy $C_{sub}$ having as root node the base class 
  (least derived class that the dispatched object can be of) of the dispatched object. From the classes located in the sub-hierarchy consider, for the currently analyzed callsite, each virtual table $v_t$. Further, within these virtual tables $v_t$'s consider only the virtual function $v_f$ entries located at the offset used by the virtual object dispatch mechanism. Next, count all virtual functions to which these entries point to.}

\medskip\noindent \textbf{Strict sub-hierarchy. (ShrinkWrap) \cite{shrinkwrap}} 
  We formalize this CFI policy $\Psi$ as the tuple   
  $\big \langle P, F_{virt},  C, V, V_{sub}, M \big \rangle$ where the relations hold:
  (1) $V \subseteq C$,
  (2) $v_e \in V$,
  (3) $v_f \in F_{virt}$,
  (4) $v_t \in V$,
  (5) $V \subseteq C$,
  (6) $V_{sub} \subseteq V$,
  (7) $C \subseteq P$,
  (8) $c_s \in P$, and
  (9) $M \subseteq Cs_{virt} V \times V_{sub} \times F_{virt}$.
  
   \textit{\sysname's Analysis.} {For each virtual callsite $vc_s$ identify the virtual table $v_t$ type used.
 Take this virtual table $v_t$ from the base class $C$ of the dispatched object and build the virtual 
 table $v_t$ sub-hierarchy $V_{sub}$ having this virtual table $v_t$ as root node.
 From the virtual tables in this $v_t$ sub-hierarchy find the virtual function $v_f$ entries located at the offset used by the virtual object dispatch mechanism for this particular callsite $c_s$. Next, count each virtual function $v_f$, to which these virtual table entries $v_e$ point to.} Finally, after \sysname computes for each callsite the total calltarget set count, as described above for each policy, it sums up all results for each callsite to generate several statistics.  \looseness=-1

\section{Implementation} 
\label{implementation} 

\subsection{Data Collection and Aggregation}
\medskip\noindent\textbf{Collection.} \sysname collects the virtual tables of a program in the Clang front-end and pushes them through the compilation pipeline in 
order to make them available during link-time optimization (LTO). For each virtual table, \sysname collects the number of entries.
The virtual tables are analyzed and aggregated to virtual table hierarchies in a later step.
Other data such as direct/indirect callsites and function signatures
are collected during LTO. 

\medskip\noindent\textbf{Aggregation.} Next, we present the program primitives which are constructed by \sysname.
First, virtual table hierarchies are built based on the previously collected virtual table metadata within the Clang front-end. The virtual table hierarchies are used to derive relationships between the classes 
inside a program (class hierarchies), determine sub-hierarchy relationships and count, for example, how many virtual table entries (virtual functions)
a certain virtual table sub-hierarchy has. 
Second, virtual table sets are constructed for mapping callsites to legitimate class hierarchy-based virtual calltargets.
Third, callsite function types are constructed. These are composed of the number of parameters provided by a callsite, their types, and if the callsite is a void or non-void callsite. 
Lastly, function types are built. These are composed of the function name, the expected number of parameters, their types and an additional bit used to indicate if the function is a void or non-void function.

\subsection{CFI Defense Modeling}
\sysname implements a set of constraints for each modeled CFI-defense, which are defined as analysis conditions that model the behavior of each analyzed CFI-defense.
These constraints are particular for each CFI-defense and operate on different primitives. More specifically, different constraints of a CFI-defense
are implemented inside \sysname. The steps for modeling a CFI defense are addressed by answering the five questions listed in the following.
 (1) Which of \sysname's primitives are used by the policy?
 (2) Is there a nesting or subset relation between these primitives?
 (3) Does the policy rely on hierarchical metadata primitives?
 (4) What are the callsite/calltarget matching criteria?
 (5) How to count a callsite/calltarget match?
Note that there is no effort needed to port \sysname from one policy to another as all policies can 
operate in parallel during compile time. As such, the measurement results obtained for each policy are written 
in one pass in an external file for later analysis.

Next, we provide a concrete example of how a CFI defense, \textit{i.e.}, TypeArmor's \textit{Bin types} policy \cite{veen:typearmor}, was modeled inside \sysname by following the steps mentioned above. For more details, see Section \ref{typearmor:policy} for a description on how this policy works.
More specifically, for TypeArmor the following applies.
 (1) The policy uses the callsite, indirect callsite, callsite function type, and function type primitives provided by \sysname.
 (2) From all functions contained in the program, we analyze only the virtual functions which expect up to six parameters to be passed by the callsite. 
 Next, from all callsites, we filter out the ones which are not calling virtual functions and which provide more than six parameters to the calltarget. 
 We check if the callsite is a void or non-void callsite. Further, we check if each
 analyzed calltarget is a void or non-void target.
 (3) The policy does not rely on hierarchical metadata.
 (4) A callsite matches a calltarget if it provides less or the same number of parameters as the calltarget expects.
 (5) In case the matching criteria holds, we increment the total count by one for each found match. \looseness=-1

Lastly, these constraints are implemented as a LLVM compiler module pass performed during LTO. Thus, even with limited knowledge constraints of an CFI policy can be modeled by observing how other existing policies were implemented inside \sysname.

\subsection{CFI Defense Analysis}
\sysname performs for each implemented CFI defense a different analysis. Each defense analysis consists of one or more iterations through the program primitives which are relevant for 
the CFI defense currently being analyzed. 
Depending on the particularities of a defense, \sysname uses different previously collected program
primitives. More specifically, class hierarchies, class sub-hierarchies, or function signatures located in the whole program or in certain class sub-hierarchy are individually analyzed.
During a CFI-defense analysis, statistics are collected w.r.t. the number of allowed calltargets per callsite taking into account the previously modeled CFI-defense.

As such, for a certain CFI defense (\textit{e.g.,} TypeArmor's CFI policy \textbf{Bin types}) it is required to determine a match between the number of 
provided parameters (up to six parameters) of each indirect callsite 
and all virtual functions present in the program (object inheritance is not taken into account) which could be the target (may consume up to six parameters) of such a callsite. 
In order to analyze this CFI defense and collect the statistics, \sysname visits all
indirect callsites 
it previously detected in the program and all virtual functions located in all previously recuperated class hierarchies.
Afterwards, each callsite is matched with potential calltargets 
(virtual functions).  
Lastly, after all virtual callsites/functions were visited, the generated information is shown to the analyst.

\subsection{Implementation Details}
We implemented \sysname as three link time optimization (LTO) passes and some code inside the Clang compiler to push metadata into the compiler's LTO. \sysname is built atop the Clang/LLVM (v.3.7.0) compiler~\cite{clang:llvm} framework infrastructure. 
The implementation of \sysname is split between the Clang compiler front-end (part of the metadata is  collected here), and several link-time passes, totaling 4.2 KLOC.
\sysname supports separate compilation by relying on the LTO mechanism built in LLVM~\cite{clang:llvm}. 
By using Clang, \sysname collects  front-end virtual tables and makes them available during LTO. 
Next, virtual table hierarchies are built which are used to model different CFI defenses.
Other \sysname primitives such as function types are constructed during LTO. 
Finally, each of the analyzed CFI defenses are separately modeled inside \sysname by using 
the previously collected primitives and aggregated data to impose the required defense constraints.

\section{Evaluation}
\label{Evaluation}

In this section, we address the following research questions (RQs).

\begin{itemize}
\item \textbf{RQ1:} What type of metrics are supported by the \sysname framework (Section \ref{metrics})?

\item \textbf{RQ2:} What is the residual attack surface of NodeJS after applying independently eight CFI defenses (Section \ref{rq1})? For answering this RQ, we performed a use case analysis focused on NodeJS.

\item \textbf{RQ3:} What score would each of the analyzed CFI defenses get (Section \ref{rq9})?

\item \textbf{RQ4:} How can \sysname be used to rank CFI policies based on the offered protection level (Section \ref{rq3})? 

\item \textbf{RQ5:} What is the residual attack surface for several real-world analyzed programs (Section \ref{rq2})?

\item \textbf{RQ6:} How can \sysname pave the way towards attack construction (Section \ref{rq7})?
\end{itemize}

\medskip\noindent\textit{\textbf{Test Programs.}} 
In our evaluation, we used the following real-world programs:
Nginx \cite{nginx} (Web server, usable also as: reverse proxy, load balancer, mail proxy and HTTP cache, v.1.13.7, \verb!C! code), 
NodeJS \cite{nodejs} (cross-platform JavaScript run-time environment, v.8.9.1, \verb!C/C++! code),
Lighttpd \cite{lighthttpd} (Web server optimized for speed-critical environments, v.1.4.48, \verb!C! code), 
Httpd \cite{httpd} (cross-platform Web server, v.2.4.29, \verb!C! code), Redis \cite{redis} (in-memory database with in-memory key-value store, v.4.0.2, \verb!C! code), 
Memcached \cite{memcached} (general-purpose distributed memory caching system, v.1.5.3, \verb!C/C++! code), 
Apache Traffic Server \cite{trafficserver} (modular, high-performance reverse proxy and forward proxy server, v.2.4.29, \verb!C/C++! code), and Chrome \cite{chrome} (Google's Web browser, v.33.01750.112, \verb!C/C++! code).

\medskip\noindent\textit{\textbf{Experimental Setup.}} The experiments were performed on an Intel i5-3470 CPU with 8GB of RAM running on the Linux Mint 18.3 OS. All experiments were
performed ten times to provide reliable values. If not otherwise stated, we modeled each of the eight CFI defenses inside \sysname according to the policy descriptions provided in \autoref{Mapping Defenses Into CFI-Assessor}.

\medskip\noindent\textit{\textbf{\sysname's CTR Analysis Capabilities.}}
\sysname can conduct different types of analysis based on several metrics as such CFI policies can be compared w.r.t. different aspects. 
In this paper, we decided to focus on the CTR metric as it is one of the simple ones and it is comparable with the AIR, fAIR, AIA and other related metrics. Further, note that CTR is an example metric which does not fit all needs.
Other supported metrics are presented in Section~\ref{metrics}.


\subsection{LLVM-CFI Supported Metrics}
\label{metrics}
In this section, we present four novel CFI metrics, which can be used within \sysname in order to perform different types of CFI policy-related analysis.
Note that another set of metrics was
introduced in a recent survey by Burow \textit{et al.}~\cite{cfi:survey}.
Complementing related work, \sysname helps to provide precise and reproducible measurement results when performing CFI-related investigations. This section introduces several alternatives to existing CFI metrics.

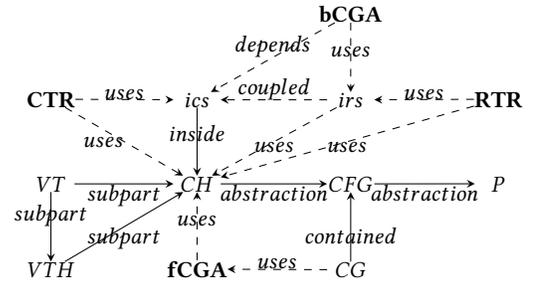
\begin{figure}[ht]
\captionsetup{justification=justified}
\centering
\tikz \node [scale = .97, inner sep = 0] {
 \begin{tikzpicture}
  \matrix (m) [matrix of math nodes,row sep=3em,column sep=4em,minimum width=2em]
  {
                    &                & \textbf{bCGA}     &                    \\
      \textbf{CTR}  & ics            & irs               &\textbf{RTR}        \\
      VT            & CH             & CFG               &P                   \\
      VTH           & \textbf{fCGA}  & CG                &                    \\};
  \path[-stealth]
    (m-2-2) edge node [above] {$inside$} (m-3-2) 
    (m-2-1) edge [dashed] node [above] {$uses$} (m-2-2) 
    (m-2-1) edge [dashed] node [left] {$uses$} (m-3-2) 
    (m-3-3) edge node [below] {$abstraction$} (m-3-4) 
    (m-3-1) edge node [above] {$subpart$} (m-4-1) 
    (m-4-1) edge node [below] {$subpart$} (m-3-2) 
    (m-3-1.east|-m-3-2) edge node [below] {$subpart$} (m-3-2) 
    (m-3-2.east|-m-3-3) edge node [below] {$abstraction$} (m-3-3) 
    (m-4-3) edge node [below] {$contained$} (m-3-3) 
    
    (m-4-2) edge [dashed] node [above] {$uses$} (m-3-2) 
    (m-4-3) edge [dashed] node [above] {$uses$} (m-4-2) 
    
    (m-2-4) edge [dashed] node [above] {$uses$} (m-2-3) 
    (m-2-3) edge [dashed] node [below] {$uses$} (m-3-2) 
    
    (m-2-3) edge [dashed] node [above] {$coupled$} (m-2-2) 
    (m-1-3) edge [dashed] node [above] {$depends$} (m-2-2) 
    (m-1-3) edge [dashed] node [above] {$uses$} (m-2-3) 
    
    (m-2-4) edge [dashed] node [below] {$uses$} (m-3-2) 
    ;
\end{tikzpicture}};
\caption{Our four metrics (bold text), \& program primitives.}
\label{Interaction of forward-edge related concepts.}
\end{figure}

\begin{table}[ht!]
\centering
\begin{tabular}{l|l}
 \textbf{Symbol} &  \ \ \ \ \ \ \ \ \ \ \ \ \ \ \ \ \ \ \ \ \ \ \textbf{Description}  \\\hline
\emph{ics} & indirect call site (\textit{i.e.,} x86 \texttt{call} instruction) \\
\emph{irs} & indirect return site (\textit{i.e.,} x86 \texttt{ret} instruction) \\
\emph{P} & program \\
\emph{VT} & virtual table \\
\emph{VTI} & virtual table inheritance \\
\emph{CH} & class hierarchy \\
\emph{CFG} & control flow graph \\
\emph{CG} & code reuse gadget \\
\emph{CTR} & indirect calltarget reduction \\
\emph{RTR} & indirect return target reduction \\
\emph{fCGA} & forward-edge based $CG$ availability \\
\emph{bCGA} & backward return-edge based $CG$ availability \\
\end{tabular}
\caption{Symbols and associated descriptions.}
\label{notation2}
\end{table}
\vspace{-.5cm}
\autoref{notation2} contains the abbreviations which we 
used in \autoref{Interaction of forward-edge related concepts.}. 

\autoref{Interaction of forward-edge related concepts.} depicts the relationships between our four metrics (bold text) metrics and program metadata primitives. 
Next, we introduce our metrics that can be used within \sysname.


\newcommand{\ctr}{\mathit{ctr}}

\begin{definition}[\textbf{CTR}]
Let $ics_{i}$ be a particular indirect callsite in a program $P$,
$\ctr_{i}$ is the total number of legitimate calltargets for an $ics_{i}$ 
after hardening a program with a certain CFI policy.
\label{ctr label}
\end{definition}
Then, the $iCTR$ metric is:
$CTR = \sum_{i=1}^n \ctr_{i}$.
Note that the lower the value of $CTR$ is for a given program, the more precise the CFI policy 
is. The optimal value of this metric is equal to the total number of callsites present 
in the hardened program. This means that there is a one-to-one mapping.
We can also capture the distribution of the numbers of calltargets
using min, max, and standard deviation functions.
Minimum: $\min_{i}{\{\ctr_{i}\}}$;
Maximum: $\max_{i}{\{\ctr_{i}\}}$; and
Standard Deviation (SD): $CTR_{SD} = \sqrt{\frac{\sum_{i=1}^n ({ctr_{i} - \overline{ctr_{i}}})^2}{n}}\label{ctrsd}$.

\begin{definition}[\textbf{RTR}]
Let $irs_{i}$ be a particular indirect return site in the program $P$, then
$rtr_{i}$ is the total number of available return targets for each $irs_{i}$ 
after hardening the backward edge of a program with a CFI policy.
\label{rtr definition.}
\end{definition}
Then, the $RTR$ metric is:
$RTR = \sum_{i=1}^n rtr_{i}$.
Note that the lower the value of $RTR$ is for a given program, the better the CFI policy 
is. The optimal value of this metric is equal to the total number of indirect return sites present 
in the hardened program. This means that there is a one-to-one mapping.
Other key properties are:
Minimum: $RTR_{MIN} = \min_{i}{\{rtr_i\}}$;
Maximum: $RTR_{MAX} = \max_{i}{\{rtr_i\}}$; and
Standard Deviation (SD): $RTR_{SD} = \sqrt{\frac{\sum_{i=1}^n {(rtr_{i} - \overline{rtr_{i}})}^2}{n}}$.

\begin{definition}[\textbf{fCGA}]
Let $cgf_{i}$ be the total number of legitimate
calltargets that are allowed and which contain gadgets according to a gadget finding tool. 
Then, the forward code reuse gadget availability $fCGA$ metric is:
$fCGA = \sum_{i=1}^n cgf_{i}$.
\label{fcga definition.}
\end{definition}
Note that the lower the value of $fCGA$ is, the better the policy is. This means that
every time a calltarget containing a code reuse gadget is protected by a CFI check, this gadget is not reachable. 
The reverse is true when the calltarget return contains a gadget and there
are indirect control flow transfers which can call this indirect return site unconstrained.

\begin{definition}[\textbf{bCGA}]
Let $cgr_{i}$ be the total number of legitimate
callee return addresses which contain code gadgets according to a gadget finding tool. 
Then, the backward code reuse gadget availability $bCGA$ metric is:
$bCGA = \sum_{i=1}^n cgr_{i}$.
\label{bcga definition.}
\end{definition}
Note that the lower the value of $bCGA$ is, the better the policy is. This means that
every time when a calltarget return address, which contains a code reuse gadget that is protected by a CFI check, then 
this gadget is not reachable. The reverse is true when the calltarget return contains a gadget and there
are indirect control flow transfers which can call this indirect return address in an unconstrained.

Important advantages of these metrics are:
(1) provide absolute numbers without averaging the results,
(2) can be used to assess backward-edge target set reduction,
and (3) can be used to assess the forward-edge and backward-edge control flow transfers w.r.t. gadget availability. 

So far, the available CFI metrics have failed to assess how likely an attack still is after a CFI policy was applied. Therefore, we presented examples of metrics that can be incorporated into our framework. 
The goal of these metrics is to point out that multiple CFI-related measurements are relevant and that these can be performed with \sysname depending on the type of CFI policy which one wants to analyze. Further, our metrics in contrast to others
can be used to analyze the protection of backward edges and the availability of gadgets after a CFI policy was deployed.

One major benefit of our metrics is that \sysname can use these to assess CFI policies w.r.t. several dimensions (\textit{e.g.,} forward-edge and backward-edge transfers, and gadget availability) and combine this information into meaningful data. Note that currently no available CFI metric can do this. Further, the security implications of these metrics 
can be used to not only tell how many targets per callsite or return site exist but also to correlate this information to gadgets which may be reachable after such an indirect transfer was executed. This can be achieved by using an additional gadget finding tool which may be used to search for CRA gadgets in order to map these within the analyzed binary. In this way, \sysname is capable of taking into account past, current and future attacks and assess their likelyhood after deploying a CFI policy.

Further, note that we did not use all four metrics in conjunction with the eight CFI policies assessed in this paper as we wanted to thoroughly focus only on forward-edge transfers in this work. Moreover, we are not aware of any CFI policy assessing tool which can take into account all the dimensions introduced by our four metrics. Lastly, by using these metrics, experiments become more reproducible and results better comparable.

\begin{table*}[ht!]
\centering
\resizebox{1.5\columnwidth}{!}{%
 \begin{tabular}{ l | r | r | r | r | r | r | r | r | r | r | r} \hline
         \multicolumn{1}{c|}{}                                          & \multicolumn{5}{c|}{\hspace{.3cm}\textbf{Targets Median}}                            & \multicolumn{6}{c}{\hspace{.2cm}\textbf{Targets Distribution}}\\
         \multicolumn{1}{l|}{\textbf{\textit{P}}}           & \multicolumn{5}{c|}{\hspace{.3cm}\textbf{}}                                          & \multicolumn{3}{c|}{\hspace{.2cm}\textbf{NodeJs}}  & \multicolumn{3}{c}{\hspace{.2cm}\textbf{MKSnapshot}}\\
         {\textbf{\textit{}}}                                 & \textit{NodeJS} & \textit{MKSnaphot} & \textit{Total} & \textit{Min} & \textit{Max}  & \textit{Min}  &\textit{90p}   &\textit{Max}   & \textit{Min}  &\textit{90p}   &\textit{Max}       \\ \hline
         \textit{(1)}                                & 21,950  & 15,817    & 15,817  & 15,817 & 21,950 & 12,545 & 30,179 & 32,478 & 8,714    & 21,785 & 23,376  \\
                  {}                                & (21,950) & (15,817)    & (20,253)  & (15,817)& (21,950) &  (885)   & (30,179) & (32,478) & (244)    & (21,785) & (23,376)  \\
         \textit{(2)}                                 & 2,885      & 2,273        & 2,273    & 2,273    & 2,885  & 0         & 5,751   & 5,751   & 1         & 4,436   & 4,436  \\
                 {}                                 & (88)      & (495)        & (139)    & (88)    & (21,950)  & (0)         & (5,751)   & (5,751)   & (0)         & (4,436)   & (4,436)  \\
         \textit{(3)}                                      & 1,511      & 1,232        & 1,232    & 1,232    & 1,511    & 0         & 5,751   & 5,751   & 1         & 4,436   & 4,436  \\
           {}                                      & (56)      & (355)        & (139)    & (56)    & (355)    & (0)         & (5,751)   & (5,751)   & (0)         & (4,436)   & (4,436)  \\
           
         \textit{(4)}                          & 3               & 2                  & 3              & 2            & 3             & 0             & 499         & 730          & 0             & 507         & 756    \\
         \textit{(5)}                                 & 6,128           & 2,903              & 6,128           & 2,903         & 6,128          & 6,128          & 6,128        & 6,128         & 2,903          & 2,903        & 2,903   \\
         \textit{(6)}                              & 2               & 1                  & 2              & 1            & 2             & 0             & 54          & 243          & 0             & 16          & 108    \\
         \textit{(7)}                                       & 2               & 1                  & 1              & 1            & 2             & 0             & 7           & 243          & 0             & 11          & 108    \\
         \textit{(8)}                        & 2               & 1                  & 1              & 1            & 2             & 0             & 6           & 243          & 0             & 9           & 108    \\
\end{tabular}
}
\caption{Legitimate calltargets per callsite for each of the eight CFI policies for NodeJS after each CFI defense was individually applied. The values not contained in round brackets are obtained for only virtual callsites and all targets (\textit{i.e.,} virtual and non-virtual), while
the values in round brackets are obtained for all indirect callsites (\textit{i.e.,} virtual and function pointer based calls) and all targets.
For the \textit{Bin types}, \textit{Safe src types}, and \textit{Src types} policies depicted  above the targets can be virtual or non-virtual, for the remaining policies the 
targets inherently can only be virt. functions.
Targets median: (min. and max.) number of legal function targets per callsite. Target distribution: minimum/90th percentile/maximum number of targets per callsite. This 90th percentile is determined by sorting the values in ascending order, and picking the value at 90\%. Thus 90\% of the sorted values have a lower or equal value to 90th percentile. P: Policy (Static target constraints), \textit{(1)} Bin types~\cite{veen:typearmor}, \textit{(2)} Safe src types~\cite{vtv:tice}, \textit{(3)} Src types~\cite{mcfi:niu}, \textit{(4)} Strict src types~\cite{zhang:vtrust}, \textit{(5)} All virtual tables~\cite{vtint:zhang}, \textit{(6)} virtual Table hierarchy~\cite{marx:tool}, \textit{(7)} Sub-hierarchy~\cite{ivt}, 
and \textit{(8)} Strict sub-hierarchy~\cite{shrinkwrap}.
}
\label{Static target constraints.}
\end{table*}
\begin{table*}[ht]
\centering
\resizebox{1.9\columnwidth}{!}{
  \begin{tabular}{ r|r|p{.3cm}| p{.9cm} | p{.3cm}|r|r|r|r|r|r|r|r } \hline
                                                                 &&    \multicolumn{1}{c|}{\textbf{Callsites}} &\multicolumn{2}{c|}{\textbf{Targets Baseline}}    & \multicolumn{8}{c}{\textbf{Virtual Function Targets}} \\
         \textbf{\textit{P}}                               &\textit{Value}       &\textit{Write cons.}  &Base all fun     &Base vFunc & \colorbox{white}{\textit{(1)}} & \colorbox{white}{\textit{(2)}} & \colorbox{white}{\textit{(3)}} & \colorbox{white}{\textit{(4)}} & \colorbox{white}{\textit{(5)}} & \colorbox{white}{\textit{(6)}} & \colorbox{white}{\textit{(7)}} & \colorbox{white}{\textit{(8)}}          \\
\hline

	        &Min	&	&	&	&12,545 (1,956)	&0 (0)	        &0 (0)	        &0 (0)	        &6,128	&0	&0	&0	\\
	        &90p	&	&	&	&30,179 (4,078)	&5,751 (810)	&5,751 (810)	&499 (10)	&6,128	&54	&7	&6	\\
JS	        &Max	&none	&32,478	&6,300	&32,478 (4,455)	&5,751 (810)	&5,751 (810)	&730 (243)	&6,128	&243	&243	&243	\\
	        &Med	&	&	&	&21,950 (3,106)	&2,885 (426)	&1,511 (121)	&3 (3)	        &6,128	&2	&2	&2	\\
	        &Avg	&	&	&	&19,395 (2,793)	&2,406 (414)	&2,113 (354)	&86 (12)	&6,128	&14	&8	&8	\\

\hline

	        &Min	&	&	&	&2,608 (232)	 &1 (0)	        &1 (0)	        &0 (0)	        &788	&0	&0	&0	\\
	        &90p	&	&	&	&4,085 (546)	 &1,315 (97)	&1,315 (97)	&17 (13)	&788	&34	&7	&7	\\
TS	&Max	&none	&6,201	&796    &6,201 (710)	 &1,315 (159)	&1,315 (159)	&18 (16)        &788	&42	&18	&18	\\
	        &Med	&	&	&	&2,608 (232)	 &1,315 (97)	&1,315 (97)	&17 (13)	&788	&7	&1	&1	\\
	        &Avg	&	&	&	&3,122 (321)     &928 (76)	&923 (74)	&11 (9)	        &788	&10	&3	&3	\\   
	
\hline

	        &Min	&	&	&	&97,041 (37,873)	&0 (0)	             &0 (0)	         &0 (0)	        &68,560	&0	&0	&0	\\
	        &90p	&	&	&	&201,477 (63,816)	&64,315 (24,661)       &64,315 (24,661)	 &48 (30)	&68,560	&192	&25	&15	\\
C        &Max	&none	&232,593	&78,992	&232,593 (71,000)	&64,315 (24,661)	     &64,315 (24,661)	 &3,029 (509)	&68,560  &4,486	&4,486	&4,486	\\
	        &Med	&	&	&	&97,041 (37,873)	&8,672 (4,593)         &7,633 (4,593)	 &3 (2)	        &68,560	&6	&2	&2	\\
	        &Avg	&	&	&	&128,452 (45,731)	&29,315 (11,119)       &29,127 (11,013)      &57 (19)	&68,560	&78	&37	&32	\\

\end{tabular}
}
\caption{Legitmate calltargets per callsite for only virtual callsites and for only the \texttt{C++} programs after each CFI defense was individually applied.
\textit{Baseline all func.} represents the total number of functions,
while \textit{Baseline virtual func.} represents the number of virtual functions.
The first four policies, from left to right in italic font (\textit{Bin types}, \textit{Safe src types}, \textit{Src types}, and \textit{Strict src types})
allow virtual or non-virtual targets, while the remaining four policies 
inherently allow only virtual targets. This is not a limitation of \sysname but rather how these were intended, designed and used in the original tools from where these are stemming.
The values in round brackets show the theoretical results after adapting the 
first four policies to only allow virtual targets.
Each table entry contains five aggregate values: minimal, 90th percentile: minimum/90th percentile/maximum,
maximal, median and average (Avg) number of targets per callsite.
P: program, JS: NodeJS, TS: Traffic Server, C: Chrome, \textit{(1)} - \textit{(8)} see \autoref{Static target constraints.}.
}
\label{Overall results vcals}
\end{table*}

\begin{table*}[ht!]
\centering
\resizebox{2.05\columnwidth}{!}{%
 \begin{tabular}{ r | r | r | r | r | r | r | r | r | r | r | r | r | r | r | r | r | r | r | r | r | r | r | r | r} \hline
         \multicolumn{1}{c|}{\textbf{P}}                                        & \multicolumn{3}{c|}{\colorbox{white}{\textit{(1)}}}     & \multicolumn{3}{c|}{\colorbox{white}{\textit{(2)}}}   & \multicolumn{3}{c|}{\colorbox{white}{\textit{(3)}}}       & \multicolumn{3}{c|}{\colorbox{white}{\textit{(4)}}}  & \multicolumn{3}{c|}{\colorbox{white}{\textit{(5)}}}     & \multicolumn{3}{c|}{\colorbox{white}{\textit{(6)}}}  & \multicolumn{3}{c|}{\colorbox{white}{\textit{(7)}}}   & \multicolumn{3}{c}{\colorbox{white}{\textit{(8)}}}  \\
         \textbf{}                                                & \textit{Avg}& \textit{SD}   & \textit{90p}    & \textit{Avg}& \textit{SD} & \textit{90p}      & \textit{Avg}     & \textit{SD} & \textit{90p}& \textit{Avg}     & \textit{SD} & \textit{90p}   & \textit{Avg}     & \textit{SD} & \textit{90p} & \textit{Avg}      & \textit{SD} & \textit{90p}& \textit{Avg} & \textit{SD} & \textit{90p}    & \textit{Avg} & \textit{SD} & \textit{90p}  \\ \hline
JS	&59.72	&21.0	&92.92	&7.41	&6.32	&17.71	&6.51	&6.44	&17.71	&0.26	&0.54	&1.54	&97.27	&0.0	&97.27	&0.23	&0.63	&0.86	&0.13	&0.46	&0.11	&0.13	&0.46	&0.1	\\
TS	&50.35	&15.79	&65.88	&14.97	&8.89	&21.21	&14.89	&9.01	&21.21	&0.18	&0.12	&0.27	&98.99	&0.0	&98.99	&1.26	&1.27	&4.27	&0.34	&0.51	&0.88	&0.34	&0.51	&0.88	\\
C	&55.23	&19.08	&86.62	&12.6	&12.16	&27.65	&12.52	&12.22	&27.65	&0.02	&0.11	&0.02	&86.79	&0.0	&86.79	&0.1	&0.43	&0.24	&0.05	&0.41	&0.03	&0.04	&0.41	&0.02	\\\hline
\textit{Avg}  &\colorbox{light-gray}{55.1}   &18.62  &81.8   &\colorbox{light-gray}{11.66}  &9.12   &22.19  &\colorbox{light-gray}{11.3}   &9.22   &22.19  &\colorbox{light-gray}{0.15}   &0.25   &0.61   &\colorbox{light-gray}{94.35}  &0.0    &94.35  &\colorbox{light-gray}{0.53}   &0.77   &1.79   &\colorbox{light-gray}{0.17}   &0.46   &0.34   &\colorbox{light-gray}{0.17}   &0.46   &0.33   \\
\end{tabular}
}

\caption{Normalized results with the baseline (B) using only virtual callsites. Note that virtual callsites can be used for all eight assesses CFI policies as these were designed in the original papers to be used for these types of callsites as well.
Baseline: Total number of possible virtual targets.
Each entry contains three aggregate values: average-, standard deviation (SD)
 and 90th percentile number of targets per callsite. The lower the \textit{Average} value is the better the CFI defense is. P: program, JS: NodeJS (Baseline 6.3K), TS: Traffic Server (Baseline 796), C: Chrome (Baseline 78,992).}
\label{Standard dev. for virtual tagerts}
\end{table*}

\subsection{NodeJS Use Case}
\label{rq1}
In this section, we analyze the residual attack surface after each of the eight CFI policies 
was applied individually to NodeJS. Note that three out of the eight assessed CFI policies used in the following tables are the same 
as reported by Veen \textit{et al.} \cite{newton} (we share the same names). For the other five CFI policies, we use names which reflect their particularities.
We selected NodeJS as it is a very popular real-world application and it contains both \verb!C! and \verb!C++! code. As such, \sysname
can collect results for the \verb!C! and \verb!C++! related CFI polices.

\autoref{Static target constraints.} depicts the static target constraints for the NodeJS 
program under different static CFI calltarget constraining policies. Further, 
\autoref{Static target constraints.} provides the minimum and
maximum values of virtual calltargets which are available for a virtual callsite after one of the eight 
CFI policies is applied. MKSnapShot contains the Chrome V8 engine and is
used as a shared library by NodeJS after compilation. We decided to add MKSnapshot in \autoref{Static target constraints.} as this component 
is used considerably by NodeJS and represents a source of potential calltargets. The NodeJS results were obtained after static linking of MKSnaphot. Further, the target median entries in \autoref{Static target constraints.} (left hand side) indicate the median values obtained for independently
applying one of the eight CFI policies to NodeJS. For both NodeJS and MKSnaphot, the best median number 
of residual targets is obtained using the following policies:
(1) \textit{vTable hierarchy}, 
(2) \textit{sub-hierarchy}, and
(3) \textit{strict sub-hierarchy}. The results indicate that these three CFI policies provide the lowest attack surface, while the highest attack surface is obtained for the \textit{bin types} policy, which allows the highest number of virtual and non-virtual targets. 

The targets distribution in \autoref{Static target constraints.} (right hand side) shows the minimum, maximum and 90 percentile results
for the same eight policies as before. While the minimum value is
0, the highest values for both NodeJS and MKSnapshot are obtained for the \textit{bin types} policy, while the lowest values are obtained for the following policies:
(1) \textit{vTable hierarchy}, 
(2) \textit{sub-hierarchy}, and
(3) \textit{strict sub-hierarchy}.
Further, the 90th percentile results show that on the tail end of the distribution, a noticeable difference between the three previously mentioned 
policies exists. We can observe that for these critical callsites the \textit{strict sub-hierarchy} policy provides the least amount of residual targets and therefore the best protection against CRAs. Meanwhile, the 90th percentile results for the \textit{strict src type} 
and \textit{vTable hiearchy} policies indicate that the residual attack surface might still be sufficient for the attacker.

\subsection{CFI Defenses Scores}
\label{rq9}
\begin{figure}[ht]
\centering
  \includegraphics[width=0.99\columnwidth, bb = 0 0 216 132]{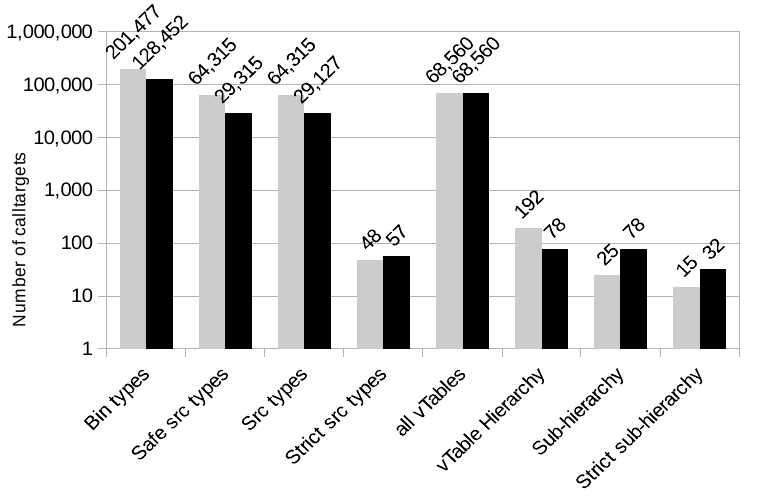}
  \vspace{-.6cm}
	\caption{Scores obtained by each analyzed CFI defense.}
	\label{scores}
\end{figure}

\autoref{scores} depicts the scores obtained by each of the eight policies, which were analyzed for the Chrome Web browser. The scores are depicted in
logarithmic scale in order to better compare the values against each other. The optimal score has the value of one depicted on the left hand side $Y$ axis. We opted to depict the values for only the Chrome browser since this represents the largest (approx. 10 million LOC) analyzed program. The numbers on the gray shaded bars represent the 90th percentile values, while the values on the black shaded bars represent the average values for the Chrome Web browser. These values are reported in \autoref{Overall results vcals} on the last row from top to bottom for the Chrome browser as well. The optimal score is one and means that each callsite is allowed to target a single calltarget. This is the case only during runtime. The lower the bar is or the closer the value is to one, the better the score is.

The best score w.r.t. the 90th percentile and average values is obtained for the \textit{strict sub-hierachy}, which appears to be the best CFI defense from the eight analyzed policies. It is interesting to note that the best function signature based policy \textit{strict src types} has a slightly worse score than the second class based CFI defense (\textit{sub-hierarchy}). Lastly, note that these CFI-based forward-edge policies are not optimal (\textit{i.e.}, they provide values larger than one)
and the desired goal is to develop policies, which provide one-to-one mappings similar to shadow stack based techniques.

\subsection{Ranking of CFI Policies}
\label{rq3}
In this section, we normalize the results presented in RQ2 using the \textit{baseline} values (\textit{i.e.,} the number of possible target functions), in order to be able to compare the assessed CFI policies against each other w.r.t. calltarget reduction.
This allows \sysname to compare the analyzed CFI defenses on programs with different sizes and complexities which would not be possible otherwise.

\autoref{Standard dev. for virtual tagerts} depicts the average, standard deviation and 90th percentile results 
obtained after analyzing only virtual callsites. Unless stated otherwise, we use the $CTR$ metric. For these callsites, all eight CFI policies can be assessed. Here, we calculated the average over the three \verb!C++! programs after normalization. By considering these aggregate average values, the eight policies can be ranked (from best (smallest aggregate average) to worst (highest aggregate average)) as follows:
(1) \textit{strict src types} (0.15),
(2) \textit{strict sub-hierarchy} (0.17),
(3) \textit{sub-hierarchy} (0.17),
(4) \textit{vTable hierarchy} (0.53),
(5) \textit{src types} (11.3),
(6) \textit{safe src types} (11.66),
(7) \textit{bin types} (55.1), and 
(8) \textit{all vTables} (94.35). 

From the class hierarchy-based policies \textit{strict sub-hierarchy} performed best in all three 
aggregate results (Avg, SD and 90th percentile). In comparison, \textit{strict sub-hierarchy} performs better w.r.t. average and standard deviation but worse w.r.t. 90th percentile. The results indicate that these two policies are the most restrictive, but a clear winner in all evaluated criteria cannot be determined.

\begin{table}[t]
\centering
\resizebox{1.0\columnwidth}{!}{%
 \begin{tabular}{ r | r | r | r | r | r | r | r | r | r | r } \hline
         \multicolumn{1}{c|}{\textbf{}}                                & &\multicolumn{3}{c|}{\hspace{0.0cm}\textbf{Bin types}}           & \multicolumn{3}{c|}{\hspace{0.0cm}\textbf{Safe src types}}     & \multicolumn{3}{c}{\hspace{0.0cm}\textbf{Src types}}    \\
         \textbf{P}                                              &\textbf{B}  & \textit{Avg}& \textit{SD}&\textit{90p}     & \textit{Avg}& \textit{SD}&\textit{90p}          &\textit{Avg} & \textit{SD}&\textit{90p}    \\ \hline
a		&32,478	&64.0	&20.43	&92.92	&3.82	&5.83	&17.71	&3.38	&5.64	&17.71	\\
b	&6,201	&54.03	&18.76	&87.89	&13.54	&9.27	&21.21	&13.46	&9.36	&21.21	\\
c	&232,593	&56.83	&19.84	&86.62	&11.71	&12.11	&27.65	&11.64	&12.16	&27.65	\\
d		&1,949	&52.18	&26.5	&92.0	&2.7	&3.01	&8.21	&2.46	&3.01	&8.21	\\
e	&594	&65.25	&27.81	&97.98	&2.94	&3.18	&7.41	&2.93	&3.19	&7.41	\\
f	&225	&69.75	&7.11	&68.89	&1.0	&0.97	&0.89	&1.0	&0.97	&0.89	\\
g		&1,270	&54.91	&24.85	&92.28	&6.38	&4.56	&11.73	&6.36	&4.57	&11.73	\\
h		&2,880	&65.19	&16.51	&84.62	&1.25	&2.52	&1.88	&1.2	&2.52	&1.88	\\\hline
\textit{Avg}&34,773	&\colorbox{light-gray}{60.3}	&34.39	&87.9	&\colorbox{light-gray}{5.4}	&5.18	&12.09	&\colorbox{light-gray}{5.3}	&5.17	&12.08	\\
\end{tabular}
}
\caption{Normalized results using all indirect callsites.}
\label{Average and standard deviation for not virtual targets}
\end{table}
\autoref{Average and standard deviation for not virtual targets} depicts (similar to \autoref{Standard dev. for virtual tagerts})
normalized results with the difference that all indirect callsites (both virtual and pointer based) are analyzed. Shortnames: B: baseline, a) NodeJS, b) Apache Traffic Server,
c) Google's Chrome, d) Httpd, e) LightHttpd, f) Memcached, g) Nginx, h) Redis.
Thus, the \textit{baseline} values used for normalization include virtual and non-virtual targets. 
By taking into account the aggregate averages and the standard deviation of the three policies
in \autoref{Average and standard deviation for not virtual targets},
we can rank the policies as follows (from best to worse):
(1) \textit{src types} (Avg 5.3 and SD 5.17),
(2) \textit{safe src types} (Avg 5.4 and SD 5.18), and 
(3) \textit{bin types} (Avg 60.3 and SD 34.39).
In contrast, by considering the 90th percentile values, we conclude that 
for the most vulnerable 10\% of callsites, \textit{bin types} only restricts the target set to 
87.9\% of the unprotected target set. As such, these callsites essentially remain unprotected.
Meanwhile, the \textit{safe src type} and \textit{src type} policies restrict
to only around 12\% of the unprotected target set.

\subsection{General Results}
\label{rq2}
\begin{table}[ht]
\centering
\resizebox{.8\columnwidth}{!}{%
 \begin{tabular}{ p{.1cm} | r | p{.8cm} | p{.99cm} | r | r | r} 
                                                                 &&\multicolumn{1}{c|}{\hspace{0.1cm}\textbf{}} &    & \multicolumn{3}{c}{\textbf{Targets (Non-) \& virt. func.}}          \\
         \textbf{\textit{P}}                               &\textit{Value}       &\textit{Callsite write cons.}  &\textit{Baseline all func.}     & \textit{(1)}    & \textit{(2)} & \textit{(3)}     \\\hline

\hline

	&Min	&	&	&885	        &0	 &0	\\
	&90p	&	&	&30,179	        &5,751	 &5,751	\\
a	&Max	&none	&32,478	&32,478	        &5,751	 &5,751	\\
	&Med	&	&	&21,950	        &88	 &56	\\
	&Avg	&	&	&20,787   	&1,242   &1,099	\\

\hline

	   &Min	        &	&	&357	&0	&0	\\
	   &90p	        &	&	&5,450	&1,315	&1,315	\\
b    &Max	        &none	&6,201	&6,201	&1,315	&1,315	\\
	   &Med	&	&	&2,608	&1,315	&1,315	\\
	   &Avg	&	&	&3,350	&840	&835	\\

\hline

	 &Min	&	&	&3,612	        &0	  &0	\\
	 &90p	&	&	&201,477	&64,315	  &64,315	\\
c    &Max	&none	&232,593&232,593	&64,315	  &64,315	\\
	 &Med&	&	&97,041	        &8,672	  &7,394	\\
	 &Avg	&	&	&132,182  	&27,238   &27,074	\\

\hline

	&Min	&	&	&99	        &0	&0	\\
	&90p	&	&	&1,793	        &160	&160	\\
d	&Max	&none	&1,949	&1,915	        &160	&160	\\
	&Med	&	&	&1,070	        &18	&16	\\
	&Avg	&	&	&1,017    	&53	&48	\\

\hline

	&Min	&	&	&37	&0	&0	\\
	&90p	&	&	&582	&44	&44	\\
e   &Max	&none	&594	&582	&44	&44	\\
	&Med	&	&	&395	&6	&6	\\
	&Avg	&	&	&388	&17	&17	\\

\hline

	  &Min	  &	&	&92	&0	&0	\\
	  &90p	  &	&	&155	&2	&2	\\
f     &Max	  &none	&225	&221	&17	&17	\\
	  &Med &	&	&155	&2	&2	\\
	  &Avg	  &	&	&157	&2	&2	\\

\hline

	&Min	&	&	&422	&1	&1	\\
	&90p	&	&	&1,172	&149	&149	\\
g	&Max	&none	&1,270	&1,259	&149	&149	\\
	&Med	&	&	&719	&75	&75	\\
	&Avg	&	&	&697	&81	&81	\\

\hline

	&Min	&	&	&1,266	        &1	&1	\\
	&90p	&	&	&2,437	        &54	&54	\\
h	&Max	&none	&2,880	&2,635	        &391	&391	\\
	&Med	&	&	&1,994	        &16	&14	\\
	&Avg	&	&	&1,877   	&36	&35	\\
                                             
\end{tabular}
}
\caption{Results for virtual and pointer based callsites. }
\label{Overall results function pointer}
\end{table}

\autoref{Overall results function pointer} depicts the
results for the three policies which can provide protection for both \verb!C! and \verb!C++! programs.
Abbreviations: {P:} program, {(1)} bin types, {(2)} safe src types, {(3)} src types; a) NodeJS, b) Apache Traffic Server,
c) Google's Chrome, d) Httpd, e) LightHttpd, f) Memcached, g) Nginx, and h) Redis.
In contrast to \autoref{Overall results vcals}, 
all indirect calls are taken into account (including virtual calls).
Therefore, the targets can be virtual or non-virtual.
Intuitively, the residual attack surface grows with the size of the program. This can be observed 
by comparing the results for large (\textit{e.g.,} Chrome) with smaller (\textit{e.g.,} Memcached) programs.

In contrast, \autoref{Overall results vcals} depicts the overall results obtained after applying the eight assessed CFI policies to virtual callsites only.
The first four policies (italic font)
cannot differentiate between virtual and non-virtual calltargets. 
Therefore, for these policies 
the baseline of possible calltargets includes all functions 
(both virtual and non-virtual). This is denoted with \textit{baseline all func}.
Since the remaining four policies can only be applied to virtual callsites, they restrict the possible calltargets to only virtual functions.
Thus, the baseline for these policies includes only virtual functions (\textit{baseline virtual function}). 
For a better comparison between the first and second categories of policies,
we also calculated the target set when restricting the first four policies
to only allow virtual callsites.
For \textit{bin types}, \textit{safe src types}, \textit{src types}, and \textit{all vTables} the results indicate that there is \textit{no} protection offered.
The three-class hierarchy-based policies perform best when considering the median and average results. In addition, the 
\textit{strict src type} policy performs surprisingly well, especially after restricting the target set to only virtual functions.

\subsection{Towards Automated CRA Construction}
\label{rq7}
In this section, we show how \sysname is used to automate one step of a COOP-like attack, namely finding protected targets which can be legitimately called.
This attack bypasses a state-of-the-art CFI policy-based defense, namely VTV's \textit{sub-hierarchy policy}.
This case study is architecture independent, since \sysname's analysis is performed at the IR level during LTO time in LLVM.
Note that LLVM IR code represents a higher level representation of machine code (metadata), 
thus our results can be applied to other architectures (\textit{e.g.,} ARM) as well.
Our case study assumes an ideal implementation of VTV/IFCC. Breaking
the ideal instrumentation shows that the defense can be bypassed in any
implementation. More specifically, we 
present the required
components for a COOP attack by studying the original COOP attack  \cite{coop} against the Firefox Web browser and 
demonstrate
that such an attack is easier to perform when using \sysname. 

For example, the original COOP attack presented by Schuster \textit{et al.} \cite{coop} consists of the following four steps:
(1) a buffer overflow filled with six fake counterfeit objects by the attacker,
(2) precise knowledge of the Firefox \texttt{libxul.so} shared library layout,
(3) knowledge about a COOP dispatcher and other gadgets (\texttt{ML-G}) resides in \texttt{libxul.so}, and
(4) how to pass information from one gadget to the other in order to open a Unix shell.
As demonstrated by COOP, the attacker first needs to find an exploitable memory corruption (\textit{e.g.,} buffer overflow, etc.) and fill it with fake objects.
Next, the attacker calls different gadgets (virtual \texttt{C++} functions) located in \texttt{libxul.so}. Note that these functions would be in the benign execution not callable as these reside in distinct class hierarchies. Further, with fine-grained CFI defenses in place these calltargets would be protected during an attack.
\sysname helps with identifying the protected targets (see step (3) above) and if desired the attacker can use other targets depending on his goals and the type of deployed CFI defense. 

As such, we assume that NodeJS contains an exploitable memory vulnerability (\textit{i.e.,} buffer overflow), and that the attacker 
is aware of the layout of the program binary. Next, we assume that the attacker wants to bend the control flow to only
per callsite legitimate calltargets since in this way he can bypass the in-place CFI policy. Next, the attacker wants to avoid calling targets located in other program class hierarchies or protected targets. Therefore, he needs to know which calltargets are legitimate for each callsite located in the main NodeJS binary and which targets are protected.

\begin{table}[ht]
\centering
\resizebox{\columnwidth}{!}{%
 \begin{tabular}{ p{.2cm}  p{.1cm}  p{.7cm}  p{.7cm}  p{.7cm}| p{.5cm}| p{.5cm}| p{.3cm}| p{.6cm}| p{.2cm}| p{.2cm}| p{.2cm}}
         \multicolumn{1}{l}{} &\multicolumn{1}{l}{}  &\multicolumn{1}{c}{} &\multicolumn{1}{c}{}                         & \multicolumn{8}{c}{\hspace{0cm}\textbf{Eight Target Policies}} \\
         \textbf{\hspace{1cm}{\textbf{CS}}}  & \textbf{\hspace{1cm}{\textbf{\#}}}         &\textbf{Base only vFunc}    &\textbf{Base all func}          & \vspace{.1cm}{\textit{(1)}}  & \vspace{.1cm}{\textit{(2)}} & \vspace{.1cm}{\textit{(3)}} & \vspace{.1cm}{\textit{(4)}} & \vspace{.1cm}{\textit{(5)}}  & \vspace{.1cm}{\textit{(6)}}        & \vspace{.1cm}{\textit{(7)}}  &\vspace{.1cm}{\textit{(8)}}    \\ \hline
         a                     & 5                   & 6,300                           & 32,478                              & 31,305              &4                        &4                   &1                           &6,128                  &1                                 &1                        &1                                 \\
         b                      & 2                   & 6,300                           & 32,478                              & 21,950              &719                      &719                 &49                          &6,128                  &57                                &53                       &49                                \\
         c                        & 3                   & 6,300                           & 32,478                              & 27,823              &136                      &136                 &1                           &6,128                  &1                                 &1                        &1                                 \\
         d                    & 1                   & 6,300                           & 32,478                              & 12,545              &810                      &810                 &1                           &6,128                  &72                                &12                       &12                                \\
         e                     & 1                   & 6,300                           & 32,478                              & 1,956               &810                      &810                 &1                           &6,128                  &72                                &13                       &13                                \\
         f                   & 1                   & 6,300                           & 32,478                              & 1,956               &810                      &810                 &6                           &6,128                  &20                                &19                       &19                                \\
         g                & 3                   & 6,300                           & 32,478                              & 1,956               &810                      &810                 &6                           &6,128                  &20                                &19                       &19                                \\
         h                      & 2                   & 6,300                           & 32,478                              & 3,106               &35                       &35                  &8                           &6,128                  &48                                &13                       &5                                 \\
         i                      & 2                   & 6,300                           & 32,478                              & 3,106               &2,984                    &2,984               &49                          &6,128                  &53                                &53                       &49                                \\
         j                      & 2                   & 6,300                           & 32,478                              & 3,106               &719                      &719                 &49                          &6,128                  &53                                &53                       &19                                \\
\end{tabular}
}
\caption{Ten controllable callsites \& their legitimate targets under the \textit{Sub-hierarchy} CFI defense. \#: passed parameters. CS: Ten controllable callsites, for \textit{(1)}-\textit{(8)}, see \autoref{Static target constraints.} caption.}
\label{Controllable callsites.}
\end{table}

\autoref{Controllable callsites.} depicts ten controllable callsites (in total \sysname found thousands of controllable callsites) for which 
the legitimate target set, depending on the used CFI policy (1-8), ranges from one to 31 to 305 calltargets: 
a)\Code{debugger.cpp:1329:33}, 
b)\Code{protocol.cpp:839:60}, 
c)\Code{schema.cpp:133:33}, 
d)\Code{handle_wrap.cc:127:3}, 
e)\Code{cares_wrap.cc:642:5}, 
f)\Code{node_platform.cc:25:5}, 
g)\Code{node_http2_core.h:417:5}, 
h)\Code{tls_wrap.cc:771:10}, 
i)\Code{protocol.cpp:839:60}, and 
j)\Code{protocol.cpp:836:60.}
For each calltarget, \sysname provides: file name, function name, start address and source code line number such that it can be easily 
traced back in the source code file.
The calltargets (right hand side in \autoref{Controllable callsites.} in
italic font) represent available calltargets for each of the eight assessed
policies.
Further, the information shown in \autoref{Controllable callsites.} demonstrates the usefulness of \sysname when used by an analyst. By using \sysname it can drastically reduce the time needed to search  for COOP-like protected and unprotected gadgets after a certain CFI policy was deployed. Lastly, this helps to better tailor attacks w.r.t. deployed CFI-based defenses.  



\section{Related Work}
\label{Related Work}
\subsection{Defense Assessment Metrics}
AIR~\cite{mingwei:sekar}, fAIR~\cite{vtv:tice}, and AIA~\cite{aia} metrics have limitations (see Carlini \textit{et al.}~\cite{carlini:bending}) and are currently the available CFI defense assessment metrics which can be used to compare the protection level offered by state-of-the-art CFI defenses w.r.t. only forward-edge transfers. 
These metrics provide average values which shed limited insight into the real offered protection level and thus cannot be reliably used to compare CFI-based defenses.
Most recently, ConFIRM \cite{confirm} also attempted to evaluate CFI, especially the compatibility, applicability, and relevance of CFI protections with a set of microbenchmarking suites. In contrast, \sysname is not a benchmark suite but rather a framework for modeling CFI defenses and comparing them against each other w.r.t. protection level these offer.
Burow \textit{et al.} \cite{cfi:survey} propose two metrics:
(1) a qualitative metric based on the underlying analysis provided by each of the assessed techniques, and 
(2) a quantitative metric that is the product of the number of equivalence classes (EC) and the inverse of the size of the largest class (LC).
In contrast, we propose \sysname, a CFI defense assessment framework and $CTR$, a new CFI defense assessment metric based on absolute forward-edge reduction set analysis, without averaging the results. $CTR$ provides precise measurements and facilitates comprehensive CFI defense comparison. \looseness=-1


\subsection{Static Gadget Discovery}
Wollgast \textit{et al.} \cite{multiarchitecture:wollgast} present a static multi-architecture gadget detection tool based on the analysis
of the intermediate language (IL) of VEX, which is part of the Valgrind \cite{valgrind} programming debugging framework. The tool can find a series of CFI-resistant gadgets. 
Compared to \sysname, both tools leave the gadget chain building as a manual effort. In contrast, when using \sysname, it is possible to define a specific CFI-defense policy 
and search for available gadgets, 
while the tool of Wollgast et al. specifies CFI-resistant gadgets by defining their boundaries (start and end instructions). These have to conform to some constraints 
and respect the normal program control flow of the program in order to be considered CFI resistant. These types of gadgets are more thoroughly described
by Goktas \textit{et al.} \cite{goktas:outofcontrol}, Schuster \textit{et al.} \cite{schuster:rop}.

RopDefender \cite{ropdefender}, ROPgadget \cite{ROPgadget}, and Ropper \cite{Ropper} are non-academic gadget detection tools based on binary program analysis.
These tools are used to search inside program binaries with the goal to find consecutive machine code instructions, which are similar to a previously specified set of rules that define a valid gadget. 
While allowing a fast search, these tools cannot detect defense-aware gadgets, since these tools do not model the defense applied to the program 
binary. As such, these tools cannot determine which gadgets are usable after a certain defense was applied. 

\subsection{Dynamic Attack Construction} 
Newton~\cite{newton}, is a runtime binary analysis tool which relies on taint analysis to help significantly simplify the detection of code reuse gadgets defined as callsite and legal calltarget pairs. 
Newton can model part of the byte memory dependencies in a given program. Newton is further able to model a series of 
code reuse defenses by not focusing on a specific attack at a time. Newton is able to craft attacks in the face of several arbitrary memory write constraints. A substantial difference compared to Newton is that \sysname uses program source code, which captures more precise information about the caller-callee pair than binary analysis based approaches.

StackDefiler \cite{losing:control:conti} presents a set of stack corruption attacks that leverage runtime object allocation information in order to bypass fine-grained CFI defenses. Based on the fact that Indirect Function-Call Checks (IFCC) \cite{vtv:tice} (also valid for VTV) spill critical pointers onto the stack, the authors show how CRAs can be built even in presence of a fine-grained CFI defense.
Compared to \sysname, which is based on control flow bending to legitimate targets, StackDefiler shows an alternative approach for crafting CRAs. More specifically, the authors show that information disclosure poses a severe threat and that shadow stacks which are not protected through memory isolation are an easy target for a skilled attacker.

ACICS~\cite{jujutsu} gadgets are detected during runtime by the ADT tool, in a similar way as Newton detects gadgets. 
Note that the ACICS gadgets are more constrained then those of Newton. For example,
only attacks where the function pointer and arguments are corruptible on the heap or in global memory are taken into consideration.
Similar to \sysname, the ADT tool is able to craft an attack in the face of IFCC's CFI defense policy by finding pairs of indirect callsites that match to certain functions 
which can be corrupted during runtime. In contrast, \sysname, is not program input dependent as it is not a runtime tool. Therefore, it can find all corruptible 
indirect callsite and function pairs under a certain modeled CFI policy.

Revery \cite{revery} crafts attacks
by analyzing a vulnerable program and by collecting runtime information on the crashing path as for example taint attributes of variables. Revery fails in some cases to generate an attack due to complicated defense mechanisms of which the tool is not aware. Lastly, in some cases, Revery cannot generate exploits due to dynamic decisions which have to be made during exploitation.

\section{Conclusion}
\label{Conclusions}

We have presented \sysname, a control-flow integrity (CFI) defense analysis framework that allows an analyst to thoroughly compare conceptual and deployed CFI defenses against each other. \sysname paves the way towards automated control-flow hijacking attack construction.
We implemented \sysname, atop of the Clang/LLVM compiler framework which offers the possibility to precisely analyze real-world programs during compile time.
We have released the source code of \sysname. By using \sysname, an analyst can drastically cut down the time needed to search for gadgets which are compatible with state-of-the-art CFI defenses contained in many real-world programs.
Our experimental results indicate that most of the CFI defenses are too permissive. Further, if an attacker does not
only rely on the program binary when searching for gadgets and has a tool such as \sysname at hand to analyze the vulnerable application, then many CFI defenses can easily be bypassed. 


\section*{Acknowledgments}
We are grateful to Elias Athanasopoulos, our shepherd, who provided highly valuable comments that significantly improved our paper. Further, we also would like to thank the anonymous reviewers for their constructive feedback. Zhiqiang Lin is partially supported by US NSF grant CNS-1834215 and ONR award N00014-17-1-2995. Gang Tan is partially supported by US NSF grant CNS-1801534 and ONR award N00014-17-1-2539.

\bibliographystyle{ACM-Reference-Format}
\bibliography{main}


\end{document}